\documentclass[preprint,12pt,authoryear]{elsarticle}
\usepackage{amssymb}
\usepackage{lineno}

\journal{Earth and Planetary Science Letters}

\begin{document}

\begin{frontmatter}

\title{The Moon turns out to be the perfect object to use the linear elasticity theory}

\author[label1]{V.P. Pavlov}
\address[label1]{Steklov Mathematical Institute of RAS, 8, Gubkina Str, Moscow, Russia.}
\ead{pavlov@mi.ras.ru}

\begin{abstract}
The applicability of the linear theory of elasticity to the Moon has been studied. As a criterion was taken
the smallness of the strain tensor. The elastic moduli are obtained from the data on the longitudinal and
transverse sound velocities in the Moon interior. The pressure was calculated in the framework of the model
of a homogeneous solid sphere under the action of its own gravity. The strain tensor trace is of the order
$0.02$, which indicates the applicability.

The equilibrium condition in the body of the Moon is considered in the reference system rigidly connected
to the rotating Moon. Except the elastic forces, all the mass forces in the body are potential. It means
that acceleration of each of them (gravity, the Earth's tidal and solar, centrifugal and inertia forces
associated with the precession of the rotation axis) is minus gradient of the corresponding potential.

It turns out that there is a hierarchy among these potentials. If you take the order of gravity for 1, the
relative order of the Earth's tide is $\gamma \sim 10^{-5}$, the Sun's one is 20 times smaller, and the
rest - less than $10^{-8}$. This allows you to keep in equilibrium condition only the Earth's tidal
potential and apply the perturbation theory on $\gamma$.

The strain tensor in the body of the Moon is calculated in the first two (zero and first) order in $\gamma$
(i.e. taking into account the Earth's tidal potential). Respectively the free energy density is calculated.
Since the axis of the Moon rotation has its own non-zero declination to the ecliptic plane, the tidal
potential variations take place during the rotation of the Moon around the Earth.

The estimation of the corresponding free energy density variations are made. Their dependence on the depth
exhibit qualitative agreement with the data of depth dependence in the energy of deep moonquakes obtained
in the project "Apollo."

Integral estimations of variations in the free energy for the year shows that it is many orders of magnitude
greater than estimates of energy for the year of deep moonquakes energy. Thus offering an answer to the
fundamental question: where is the source of energy released in the deep moonquakes.
\end{abstract}


\end{frontmatter}


\section{Introduction.}
\label{1}

Any statements on the planets seismicity are based on the accumulated information about their internal
structure. To a large extent, this information is obtained by processing of data on seismic events. First
of all, these are the data on the depth dependence of the longitudinal $c_p$ and transverse $c_s$ speed of
sound in a solid body of a planet. Here one should note three circumstances. First, the speed values are
reconstructed within linear elasticity theory. Second, the conventional relation between elastic moduli
(bulk elastic modulus $K$ and shear modulus $\mu$) and speeds $c_p$ and $c_s$ and density $\varrho$
\begin{eqnarray}\label{1}
\mu = \varrho c_s^2, \quad K = \varrho(c_p^2 - \frac{4}{3}c_s^2)
\end{eqnarray}
holds only when the coefficients in the wave equation describing the propagation of sound are constant, i.e.
when all the above mentioned parameters of the medium are constant. Third, there is a natural applicability
criterion of elasticity linear theory, the smallness of the strain tensor $u_{ik}$. In this theory, the trace
of the strain tensor $u_{ii} = p/K$, where $p$ is pressure. Let us take as a criterion of applicability of the
smallness of the trace and see how it is met in geophysical models.

We note first of all that in those models (e.g., \citep{Bul,Stac,Gar})  is used for the pressure the
Bullen-Haddon hydrostatic approximation ($\mu = 0$) in which the pressure at the surface is zero. In general,
this is not so: body forces create internal stresses (including shear) even in the absence of external
forces. In particular, the model of a homogeneous gravitating sphere (exercise 3 of \S7 in \citep{Land})
gives
\begin{eqnarray}\label{2}
p(r) = p(0)\Bigl(1 - \Bigl(1+\frac{8\mu}{15K}\Bigr)^{-1}\Bigl(\frac{r}{R}\Bigr)^2
\Bigr)\Bigr)
\end{eqnarray}
Accordingly, for estimation of the criterion we use the model (2) for the pressure and the formulas (1) to the
elastic moduli.

Conventional data (see, for example, \citep{Stac}) on the mechanical parameters of the Earth give $8\mu/15K \sim
0.27$ for the whole mantle. For the pressure is reasonably to use normalization to the common value of the
pressure at the lower boundary of the mantle. Then there are the following values $u_{ii}$ at depths of 60, 800
and 2878 $km$: 0.19, 0.15 and 0.21. It means that with such precision linear elasticity, in framework of which
the mechanical parameters and evaluated, is applicable. (Note that four significant figures for them are issued
everywhere in conventional tables!)

For the Moon (\citep{Gar}) $8\mu/15K \sim 0.3$, the value $p(0)$ is calculated in \citep{Pav}:
\begin{eqnarray}\label{3}
  p(0) = \frac{2}{3}\pi\kappa\varrho^2 R^2 \Bigl(1+\frac{4\mu}{3K}\Bigr)^{-1}\Bigl(1 +\frac{8\mu}{15K}\Bigr) \sim 29\times10^9 din\, cm^{-2}
\end{eqnarray}
and estimates for the values of $u_{ii}$ at the depth of $60, 270, 400$ and $800\, km$ are as follows:  $0.008,
0.011, 0.012$ and $0.017$ respectively. Estimation errors are relates only to errors in the measurement of
sound velocity. For $c_p$ they are within $3$\% throughout the entire interior of the Moon, and for $c_s$ they
increase to $11$\% only on the boundary of the mantle and the liquid core. Moreover, with $7$\% accuracy
 density and elastic modules can be considered constant throughout the thickness of the mantle since the depths
 of$28 km$ to the border with the liquid core. Thus, unlike the Earth, the Moon is a perfect object to use
 linear elasticity.

\section{The Moon model.}
\label{2}
Let us apply the linear elasticity theory to describe the variations of the Moon interior stress state. We will
work in a frame of reference rigidly attached to the rotating Moon. In this system, all the mechanical forces
(except elastic) acting in the Moon interior (gravity, inertia forces and tidal) are potential: the acceleration
of each of them is the negative gradient of the corresponding potential.

Since the polar flattening of the Moon is extremely small ($1.25\times10^{-3}$), for the gravitational potential
 $\varphi_0$ it is enough to take the approximation of a homogeneous sphere
\begin{eqnarray}\label{4}
 \varphi_0(r) = \frac{2}{3}\pi\kappa\varrho R^2 \Bigl(3 - \Bigl(\frac{r}{R}\Bigr)^2\Bigr),
\end{eqnarray}
where $\kappa=6.67\times 10^{-8}cm^3 g^{-1}s^{-2}$ is the gravitational constant, $\varrho=3.34 g\,cm^{-3}$ the
average density of the Moon, $R=1.737\times 10^{8}cm$ is its radius, $r$ is the length of the radius vector
$\emph{\textbf{r}}$ of point inside the Moon. The order of $\varphi_0$ is determined by the coefficients in
front of the bracket (4) and is $1.4\times 10^{10} cm^2 s^{-2}.$

Motion of the Moon in the solar system is rather complicated. Working in our frame of reference, we must
consider the potentials of all the inertial forces associated with the acceleration of this movement. We are
interested the dependence on $\emph{\textbf{r}}$ of the forces generated by these potentials.

Centrifugal potential associated with the proper rotation of the Moon, expressed by the formula
\begin{eqnarray}\label{5}
\varphi_{cf}(\textbf{\emph{r}}) = \frac{1}{2}(\omega_j ^2 r^2- (\omega_j r_j)^2),
\end{eqnarray}
where $\omega_j$ is angular velocity vector with the value of $1.75\times 10^{-7} sec^{-1}$ and an inclination
relative to the ecliptic plane $\epsilon = 1.54^{\circ}$. The order of this potential we estimate as its
magnitude for $r = R$ and obtain $4.6\times 10^2 cm^2 s^{-2}$.

Centrifugal potential $\varphi_{cf(1)}$ associated with the rotation of the Moon around the Earth, expressed
 by the similar to (5) formula. It contains instead of $\omega$ the angular velocity $\omega_1$ of the Moon
 orbital motion around the Earth. The magnitude of $\omega_1$ is the same as of $\omega$, the average
 inclination $i$ with respect to the ecliptic plane equals $5.145^\circ$, and the order of the
 $\varphi_{cf(1)}$ is same that of (5). A similar formula holds for the centrifugal potential
 $\varphi_{cf(2)}$, associated with the rotation of the Earth-Moon pair around the Sun. The angular velocity
 $\omega_2$ has the magnitude of $1.5\times 10^{-8} s^{-1}$,  $\omega_2$ is perpendicular to the plane of the
 ecliptic; the order of $\varphi_{cf(2)}$ equals to $3.4 cm^2 s^{-2}$.

The last of the force of inertia potential $\varphi_{pr}$ is associated with the precession of the lunar orbit
 axis with a period of $18.6\, years = 7.8\times 10^9 s$. Its angular velocity vector $\varphi_{pr}$ has the
 magnitude $0.8\times 10^{-9} s^{-1}$ and deviates from the direction of $\omega_1$ at an angle $i$.
 $\varphi_{pr}$ order is a hundred times less than that of $\varphi_{cf(1)}$.	

For the tidal potentials of the Earth and the Sun is sufficient to take the Laplace approximation
\begin{eqnarray}\label{6}
 \varphi_1(\textbf{\emph{r}}) = \frac{1}{2}\kappa \tilde{M}\tilde{R}^{-3}R^2 \Bigl(\frac{r^2}{R^2}-3\frac{(\textbf{\emph{r}}
 \tilde{\textbf{\emph{R}}})^2}{R^2 \tilde{R}^2}\Bigr),
\end{eqnarray}
where $\tilde{M}$ and $\tilde{\textbf{\emph{R}}}$ are the mass of the tide generating body and distance to it
 from the Moon center. Orders of these two potentials are defined as for (4), by the coefficients before of
 the bracket; for the Earth the order is $1.1\times 10^5 cm^2 s^{-2}$, and for the Sun it is $6\times
 10^3 cm^2 s^{-2}$.

We see that there is a kind of hierarchy for the potentials: if we take the gravitational potential as 1, the
Earth tidal potential has the order $0.8\times 10^{-5}$, the Sun one has the order $4\times 10^{-7}$, the
centrifugal potentials $\varphi_{cf}$ and $\varphi_{cf1}$ have the order $3\times 10^{-8}$, for $\varphi_{cf2}$
the order is $2.4\times 10^{-10}$. We emphasize that the time dependence is only at the tidal potential with a
characteristic period of 1 month: in our frame of reference only $\tilde{R}$ depend on time. Therefore, only
the variations of the tidal potential can cause variations in the stress state in the the Moon interior. Since
the own oscillations are damped in the Moon interior in a few hours, to estimate the tidal effects impact on
the stress state of the Moon interior, we can use the condition of equilibrium instead of the equations of
motion.

\section{The linear elasticity equations.}
\label{3}
In the linear theory of elasticity, Hooke's law is valid:
\begin{eqnarray}\label{7}
\sigma_{ik} = K\delta_{ik}u_{jj} + 2\mu(u_{ik} - 3^{-1}\delta_{ik}u_{jj}) = -p\delta_{ik} + 2\mu u'_{ik},
\end{eqnarray}
where $u_{ik} = 2^{-1}(\partial_i u_k + \partial_k u_i)$ is the strain tensor, $u_i = u_i (\emph{\textbf{r}})$
is the displacement vector, $\partial_i = \partial/\partial r_i$, $p$ is pressure and $u'_{ik}$ defined by
formula (7) is the deviator.

We write the equilibrium condition as a differential equation for the displacement vector:
\begin{eqnarray}\label{8}
\varrho\partial_i \varphi = \partial_k\sigma_{ik} = ((K+3^{-1}\mu)\partial_i \partial_k + \mu\delta_{ik}\triangle)u_k \equiv L_{ik}u_k,
\end{eqnarray}
where $\varrho$ is density, $\triangle =\partial_j^2$ is Laplace operator and $\varphi = \Sigma_a \varphi_a$
is the potentials sum of mass forces (gravity, centrifugal and tidal). $L_{ik}$ for our Moon model is the
second order differential operator with the constant coefficients. (Everywhere in previous formulas and later
the sum on repeated indices is implied: $a_k b_k \equiv \Sigma_k a_k b_k$.)

Boundary condition for a differential equation (8) for $u_i$ is set on the Moon surface $\Sigma$ defined as
$\varphi = const$:
\begin{eqnarray}\label{9}
\sigma_{ik}\partial_k\varphi|_{_\Sigma} = 0.
\end{eqnarray}

The solution of equilibrium condition for $u_i$ is the sum of a particular solution of the inhomogeneous
equation $L_{ik}u_k = \varrho\partial_i \varphi$, represented as a simple gradient, $u_i = \partial_i
u(\emph{\textbf{r}})$, and a general solution of the homogeneous equation $L_{ik}\tilde{u}_k = 0$, represented
as the sum $\tilde{u}_k = \partial_i \tilde{u}(\emph{\textbf{r}}) + w_i(\emph{\textbf{r}})$ of the gradient
and the vector $w_i$ divergence of which is zero: $\partial_i w_i = 0.$ The scalar functions $u$ and
$\tilde{u}$ naturally called deformation potentials.

The inhomogeneous equation takes the form of equation for $p = -K\triangle u$ -- the contribution to the
pressure  from the inhomogeneous equation solution:
\begin{eqnarray}\label{10}
\partial_i p = -A\varrho\partial_i\varphi, \quad A=(1+4K/3\mu).
\end{eqnarray}

The solution of equation (10) can be obtained by the following argument (\citep{Pav}). The variables $p,
\varrho$ and $\varphi$ are the scalars relative to rotation group $O(3)$ and must depend on scalar combinations
of their argument $\emph{\textbf{r}}$ only. (Here we may temporarily consider $\varrho$ as changeable.) We may
take the potentials $\varphi_0$ and two $\varphi_1$ for such scalar combinations denoting them as $\varphi_a,
a = 0, 1, 2$. Then the equations (10) take the form
\begin{eqnarray}\label{11}
(\partial_a p = A\varrho)\partial_i\varrho_a, \partial_a = \partial/\partial\varphi_a
\end{eqnarray}
for each $a = 0, 1, 2$. But the gradients $\partial\varphi_a$ are linearly independent and therefore
$\partial_a p - A\varrho = 0$.  It means that $p$ depends on whole sum $\varphi$ only and it is valid the
formula
\begin{eqnarray}\label{12}
p(\emph{\textbf{r}})=-A\int \varrho d\varphi = -A\varrho \varphi(\emph{\textbf{r}}) + p_0.
\end{eqnarray}
(In (12) we come back to the constant $\varrho$ of our Moon model.)  Note that formula (12) gives the
generalization of the above relation (2) and allows for corrections by other than the gravitational potential.
Integration constant $p_0$ is to be found from the boundary conditions (9).

The homogeneous equation has the form
\begin{eqnarray}\label{13}
\partial_i \tilde{p} - A\mu\triangle w_i =0, \quad \tilde{p} = -K\triangle \tilde{u}.
\end{eqnarray}
Take its divergence. We obtain the Laplace equation for $\tilde{p}$. The boundary condition for it can be
taken zero as the pressures $p$ and $\tilde{p}$ are included in  boundary condition (9) as summands. But
harmonic function with zero values at the boundary is equal to $0$ everywhere. Therefore $\tilde{u}$ and
$w_i$ are harmonic functions.

To use the boundary condition, we need to find the deformation potentials $u$ and $\tilde{u}$ and harmonic
vector-function $w_i$.

Having formula (12) for pressure and the relation $p = -K\Delta u$ between pressure and the trace of the
strain tensor, we can recover the deformation potential $u$ and then the stress and strain tensors, as well
as satisfy the boundary condition (9).

According to the arguments presented above, the scalar function $u$ depends only on $\varphi_0$ and $\varphi_1$.
In what follows, as the scalar arguments of the function $u$ it is convenient to take the variable part of the
gravity potential $x = \alpha_0 r^2/2$ and the dimensionless ratio $z = \varphi_1 /x$ while distinguishing the
small parameter $\gamma = \alpha_1/\alpha_0 \sim 10^{-5}$ in it: $z = \gamma\zeta,\,
\zeta = -r^{-2}\tau_{ik}r_i r_k.$ (Here we denote by $\alpha_0$ and $\alpha_1$ the coefficients before $r^2/2$
in expressions (4) and (6).) In these terms, the equation for recovering $u$ has the form
\begin{eqnarray}\label{14}
\Delta u(x,z) =B(x(1+\gamma\zeta)-c), \quad B=AK^{-1}\varrho,\quad c=\frac32\alpha_0 R^2 + (A\varrho)^{-1}p_0.
\end{eqnarray}
It is shown in \citep{Pav} that a solution to equation (14) is given by
\begin{eqnarray}\label{15}
u =\tilde{B}\Bigl(x^2\Bigl( \frac15-\frac27\gamma\zeta\Bigr)- \frac23 cx\Bigr), \quad
\tilde{B}=(2\alpha_0)^{-1}B.
\end{eqnarray}
It gives us a particular solution of the inhomogeneous equation $L_{ik}u_k = \varrho\partial_i\varphi$ for
the deformation vector $u_k = \partial_k u.$

The general solutions for harmonic functions $\tilde{u}$ and $w_i$ are constructed within the framework of
perturbation theory in the small parameter $\gamma.$ The Laplace equation for $\tilde{u}(x,z)$ is given by
\begin{eqnarray}\label{16}
\Delta\tilde{u}(x,z) = 2\alpha_0 x^{-1}\Bigl(\frac32 x u_x + x^2 u_{xx} - \frac32 \gamma\zeta u_z + \gamma^2 (\zeta^2 - \zeta +2)u_{zz}\Bigr) = 0.
\end{eqnarray}
In the first two orders in $\gamma$ the general solution for $\tilde{u}(x,z)$ with the natural condition
$\tilde{u}(0,z) < \infty$ has the form
\begin{eqnarray}\label{17}
\tilde{u} = c_1 xz,
\end{eqnarray}
where $c_1$ is an arbitrary constant. It is shown in \citep{Pav}, that the vector function $w_i$ vanishes in
the first two orders in $\gamma$.

Thus, in the first two orders in $\gamma$, the general solution of the equilibrium conditions for the
displacement vector $u_i(\textbf{r})$ is given by the sum of the gradient of the deformation potential $u$
defined by formula (15) and the gradient of the deformation potential $\tilde{u}$ defined by (17). It is shown
in \citep{Pav}, that the terms of the zeroth order in $\gamma$ of the boundary condition (9) fix the constant
$c$:
\begin{eqnarray}\label{18}
c = \frac{\alpha_0}{2}R^2\Bigl(1 + \frac{8\mu}{15K}\Bigr),
\end{eqnarray}
and the terms of the first order do it for the constant $c_1$:
\begin{eqnarray}\label{19}
c_1 = -\frac{\alpha_0}{2}R^2\frac{2}{35}.
\end{eqnarray}
As a result, formula (15) and (17) -- (19) allow to calculate the strain tensor and stress tensor up to first
order in $\gamma$. In particular, we get for the pressure
\begin{eqnarray}\label{20}
p(\textbf{r}) = \frac12 \alpha_0\varrho \Bigl(1 + \frac{4\mu}{3K}\Bigr)^{-1}\Bigl(\Bigl(1 +
\frac{8\mu}{15K}\Bigr)R^2 - r^2\Bigl(1-\gamma\tau_{ik}\frac{r_i r_k}{r^2}\Bigr)\Bigr).
\end{eqnarray}
Obviously when $\gamma = 0$ we obtain (2) for uniform solid sphere, and when $r = 0$, -- the value $p (0)$
of formula (3). It should be noted that the members of the first order in $\gamma$ in (20) do not contribute
to the value $p(0).$

\section{Variations of the free energy density}
\label{4}

In linear elasticity theory, the volume density $f$ of the free energy of elastic stresses is expressed in
terms of the stress tensor:
\begin{eqnarray}\label{21}
f = \frac{K}{2} u_{jj}^2 + \mu(u'_{ik})^2.
\end{eqnarray}

In Section \textbf{3}, we have calculated the trace of the strain tensor $u_{jj} = \Delta u$ (formula (14)).
The results of the same section allows to calculate the deviator $u'_{ik} = u_{ik} - 3^{-1}\delta_{ik}u_{jj}.$
In both formulas, only the dimensionless function $\zeta = 1 - 3 \cos\vartheta$ depends on time, where
$\vartheta$ is the angle between the radius vector $\emph{\textbf{r}}$ of a point in the Moon's body and
the radius vector $\tilde{\textbf{\emph{R}}}$ of the Earth.

The function $\zeta$ depends periodically (with a period of $T$ = 1 month) on time and experiences variations
$\delta\zeta$. Define $\delta\zeta$ as the difference of absolute values of $\zeta$ for two extreme positions
of $\tilde{\textbf{\emph{R}}}$ in our coordinate system. In the spherical coordinates, $\emph{\textbf{r}} =
r(\cos\theta\cos\phi, \cos\theta\sin\phi, \sin\theta)$, where $\theta$ and $\phi$ are the latitude and
longitude of the radius vector $\emph{\textbf{r}}$ in the Moon's body (the axis $z$ is directed along the
self-rotation axis of the Moon). Since the self-rotation of the Moon and its motion along the orbit are
synchronized and the direction of the angular velocity vector $\omega_j$ is constant, from the
standpoint of the Moon the extreme positions of the Earth correspond to $\tilde{\textbf{\emph{R}}} =
R(\cos\varepsilon, 0,\pm\sin\varepsilon)$, where $\varepsilon$ is the angle of inclination of $\omega_j$
to the ecliptic plane. Corrections related to the ellipticity of the Moon's orbit for the
$\emph{\textbf{r}}$-dependent part of $\zeta$ are on the order of $0.05,$ and we neglect them. As a result,
we obtain the following formula for the Earth tides:
\begin{eqnarray}\label{22}
\delta\zeta = |\sin 2\varepsilon|\cdot |\sin 2\theta|\cdot |\cos\phi|.
\end{eqnarray}

The variation of the function $\zeta$ corresponds to the variation of the free energy density (21):
\begin{equation}\label{23}
\delta f = Ku_{0jj} \delta u_{1jj} + 2\mu u'_{0ik}\delta u'_{1ik},
\end{equation}
where $u_{0jj} = B(x - c)$ is the trace of the strain tensor in the zeroth order in $\gamma$, $\delta u_{1jj}
= B\gamma x\delta\zeta$ is the variation of the trace in the first order in $\gamma$, and the convolution
$u'_{0ik}\delta u'_{1ik}$ is calculated in paper \citep{Pav}:
\begin{equation}\label{24}
u'_{0ik}\delta u'_{1ik} = B^2\gamma\frac{22}{105}x\Bigl(x + \frac{21}{22}c_1\Bigr)\delta\zeta.
\end{equation}

As a result, we obtain the following formula for the variations of the free energy density:
\begin{equation}\label{25}
\delta f = B^2 \gamma\delta\zeta x \Bigl(K(x-c) + \frac{44}{105}\mu\Bigl(x+ \frac{21}{22}c_1\Bigr)\Bigr).
\end{equation}
Here it is useful to introduce a new variable $y = r/R$ and substitute the mean values of elastic moduli,
$\mu/K \sim 0.57.$ Then (25) reduces to
\begin{equation}\label{26}
\delta f = - 0.75\gamma\delta\zeta K^{-1}p^2(0) y^2 (y^2 - 1.04^2).
\end{equation}
The graph of the variation of the free energy density versus the dimensionless radius $y = r/R$ has a zero at
the center, a maximum at $y = 0.52$, and one-third of the maximum value on the surface (in the hydrostatic
approximation it would be zero).

\section{Comparison with seismic data}
\label{5}

The data on moonquakes obtained in $1969-1977$ by the expeditions of the Apollo project (\citep{Nakal,Lamm})
have been discussed in the literature for more than four decades. Today, there is a catalogue
(\citep{Nak2003,Nak2005}) of more than 12 500 seismic events. Most events (more than 7000) are identified as
deep moonquakes whose sources are concentrated at about 300 "nests" at a depth of between 700 and 1200 $km$.

According to the modern views on the mechanics of strength and destructions (see, for example, \citep{Vol}),
periodic variations of pressure make the main contribution to the variations of the free energy density and
lead (in addition to ordinary dissipation into heat) to the accumulation of defects in a solid medium
(dislocations, cracks, etc.). Such a process is accompanied by the concentration of energy in these structures.
When the concentration reaches a certain critical limit, the medium is destroyed with a release of accumulated
energy.

We draw attention to the qualitative similarity between two patterns: depth distributions for deep focus
moonquakes and variations of the free energy density due to tides. This similarity suggests that it is the
energy concentrated at defects that is released in deep focus moonquakes.

This suggestion is corroborated by the integral (over the depths between 700 and 1200 $km$, which correspond
to the interval $[0.3R, 0.7R]$) estimate for the variations of the free energy, which is possible due to
formula (26):
\begin{equation}\label{27}
\delta F = \int \delta f(r)r^2drd\cos\vartheta d\varphi
\end{equation}
over a half-period $T/2$. For the Earth tides, this estimate yields $0.9 \times 10^{29} erg$ during a period of
$T/2 = 1.8 \times 10^7 s$, or $2 \times 10^{30} erg$ per year. This is mainly the energy of oscillations.
However, in a nonideal continuum, the dissipation of the energy of oscillations occurs both due to viscosity
and due to the accumulation of defects.

Available estimates for the fraction of dissipating energy are very rough, depend on how the nonideality of
the medium is modeled, and are on the order of $10^{-2}$. Apparently, the overwhelming part of this energy is
converted into heat and is spent on heating the body of the Moon. If all the heat goes outside, then the power
of its flux amounts to $1.4 \times 10^2 erg s^{-1} cm^{-2} = 14 \times 10^{-6} W cm^{-2}$. The available
experimental estimates (\citep{Lang}) yield a value of $2 \times 10^{-6} W cm^{-2}$. Hence, theoretically,
all the heat is spent on heating the body of the Moon.

The fraction of dissipating energy that is spent on the accumulation of defects is also estimated roughly
(\citep{Vol}). The data of laboratory experiments are formulated in terms of the number of cycles of periodic
loading that lead to the destruction of a sample. Translation of these data into the language of energy stored
in defects depends on the model of the destruction process and is estimated as $10^{-5}$ of the oscillation
energy. In any case, this estimate is many orders of magnitude greater than the estimate of energy release in
deep focus moonquakes (see \citet{Goi}), which amounts to $8 \times 10^{13} erg$ per year. We conclude that the
energy of tidal oscillations is more than enough to explain where the energy released in deep focus moonquakes
comes from.\newline

\section{\textbf{Acknowledgements}\newline}

I am grateful to N.A. Slavnov and to the participants of the seminar led by A.G. Kulikovskii for the discussion
of the results and valuable remarks.

This work is supported by the Russian Science Foundation under grant 14-50-00005.\newline




\end{document}